\documentclass[amsmath,amssymb,11pt,superscriptaddress,reprint, preprintnumbers, notitlepage,aps,prl,twocolumn]{revtex4-2}
\pdfoutput=1 
\usepackage[utf8]{inputenc}
\usepackage[english]{babel}
\usepackage{amsmath}
\usepackage{graphicx}
\usepackage{dcolumn}
\usepackage{pbox}
\usepackage{amssymb}
\usepackage{epsfig}
\usepackage[dvipsnames]{xcolor}
\usepackage[normalem]{ulem}
\usepackage{slashed}
\usepackage{amssymb}
\usepackage{verbatim} 
\usepackage{ mathrsfs }
\usepackage{color}
\usepackage[font=small]{caption}
\usepackage[font=small]{subcaption}
\usepackage{url}
\usepackage[export]{adjustbox}
\definecolor{MyDarkBlue}{rgb}{0.1, 0.1, 0.8}
\definecolor{SBlue}{rgb}{0.2, 0.4, 0.7} 
\definecolor{MyLightBlue}{rgb}{0.22,0.51,0.9}
\definecolor{MyGreen}{rgb}{0.0, 0.5, 0.0}
\definecolor{BrickRed}{rgb}{0.8, 0.25, 0.33}
\RequirePackage{hyperref}
\hypersetup{colorlinks, citecolor=SBlue,linkcolor=MyDarkBlue, urlcolor=PineGreen}

\newcommand{\cmmnt}[1]{\ignorespaces}

\makeatletter

\makeatletter
\renewcommand\@makecaption[2]{%
  \par
  \vskip\abovecaptionskip
  \begingroup
  
   \small\rmfamily
    \begingroup
     \samepage
     \flushing
     \let\footnote\@footnotemark@gobble
     \@make@capt@title{#1}{#2}\par
    \endgroup
  \endgroup
  \vskip\belowcaptionskip
}
\makeatother

\DeclareUnicodeCharacter{2212}{-}
\begin{document}
\title{\vspace{1cm}\large 
Tracing Neutrino Non-Standard Interactions 
\\through Charged Lepton Collisions
}

\author{\bf Sudip Jana}
\email[E-mail:]{sudip.jana@okstate.edu}
\affiliation{
Harish-Chandra Research Institute, A CI of Homi Bhabha National Institute, Chhatnag Road, Jhunsi, Prayagraj 211019, India}
\author{\bf Saumyen Kundu}
\email[E-mail:]{saumyenkundu@hri.res.in}
\affiliation{
Harish-Chandra Research Institute, A CI of Homi Bhabha National Institute, Chhatnag Road, Jhunsi, Prayagraj 211019, India}
\author{\bf Santosh Kumar Rai}
\email[E-mail:]{skrai@hri.res.in}
\affiliation{
Harish-Chandra Research Institute, A CI of Homi Bhabha National Institute, Chhatnag Road, Jhunsi, Prayagraj 211019, India}

\begin{abstract}
Neutrino non-standard interactions (NSI) play a crucial role in neutrino oscillations and can provide valuable insights for constructing models of neutrino masses and mixing.
While NSI have been widely explored through oscillation and scattering experiments, as well as in cosmological and astrophysical contexts, we focus on probing them at future lepton colliders like the ILC, CLIC, and FCC-$ee$. If NSI arise from heavy mediators above the electroweak scale, these colliders can offer superior sensitivity compared to neutrino experiments across a broad mass range. A notable outcome is that lepton collider data can help resolve parameter degeneracies seen in oscillation studies. We find that large NSI scenarios, proposed to address the tension between T2K and NO$\nu$A results, can be completely tested at such collider facilities. We also explore the potential of future colliders like the FCC to probe leptonic NSI using lepton PDFs in proton-proton collisions.
\end{abstract}
\hfill HRI-RECAPP-2025-06
\maketitle
\textbf{\emph{Introduction}.--}
Over the past several decades, a vast array of data from solar, atmospheric, accelerator, and reactor neutrino experiments has provided compelling and consistent evidence for neutrino oscillations, firmly establishing that neutrinos possess non-zero masses and exhibit flavor mixing. Despite this breakthrough, the fundamental origin of neutrino mass and mixing remains elusive. A wide range of theoretical frameworks has been proposed to address this puzzle, each introducing distinct mechanisms to account for the smallness of neutrino masses—yet none are accommodated within the Standard Model (SM). Notably, many of these models inherently predict new interactions between neutrinos and matter, known as non-standard interactions (NSIs). These NSIs are often an unavoidable consequence of mass-generation theories, though their strength and structure can vary significantly depending on the underlying mechanism or energy scale involved (for recent reviews, see Ref.~\cite{Proceedings:2019qno}). Recent developments \cite{Babu:2019mfe} have shown that neutrino mass models can be broadly classified into two categories based on the predicted magnitude of NSIs: Type-I models, which allow for sizable NSIs, and Type-II models, which predict them to be negligibly small. Given the unresolved nature of neutrino mass generation, it is imperative to maintain an open and model-independent perspective, particularly in exploring whether observable NSIs might exist and how they could be probed experimentally. The notion of NSIs was pioneered by Wolfenstein in 1978 \cite{Wolfenstein:1977ue} and has since attracted considerable theoretical and experimental attention as a vital avenue for exploring physics beyond the Standard Model in the neutrino sector.

NSIs can significantly alter the effective matter potential encountered by neutrinos as they propagate, thereby introducing notable complexity into the extraction of fundamental neutrino oscillation parameters. Historically, NSIs were considered as a potential explanation for the solar neutrino deficit before the Mikheyev--Smirnov--Wolfenstein (MSW) mechanism was firmly established as the leading solution. Nevertheless, NSIs continue to influence interpretations of solar neutrino data. For example, their presence can introduce degeneracies in determining the solar mixing angle $\theta_{12}$ \cite{Miranda:2004nb}. Moreover, NSIs have been shown to mitigate the mild tension between solar and KamLAND data by flattening the high-energy portion of the solar neutrino spectrum (above $\sim$3 MeV) and enhancing the predicted day-night asymmetry \cite{Maltoni:2015kca}. In the context of long-baseline accelerator experiments, sizeable flavor-changing NSIs have also been suggested as a potential solution of the observed differences in the extracted values of the CP-violating phase $\delta_{\mathrm{CP}}$ between NO$\nu$A and T2K \cite{Denton:2020uda, Chatterjee:2020kkm}. Thus, NSIs play a pivotal role in shaping our understanding of neutrino oscillation phenomena. Numerous studies have explored their implications in oscillation and scattering experiments, yet their effects remain relatively underexplored at the energy frontier, where only a limited number of investigations have been conducted to date.

Collider experiments offer several distinctive advantages in probing NSIs. Unlike neutrino oscillation studies, which suffer from parameter degeneracies and are insensitive to the mediator mass, collider sensitivities strongly depend on the mediator’s mass but remain unaffected by such degeneracies. Additionally, while oscillation experiments are blind to axial-vector contributions and only sensitive to differences in matter potential, high-energy collider data can independently constrain all NSI parameters, including both vector and axial components. This complementarity highlights the synergy between collider searches and oscillation data in exploring physics beyond the Standard Model in the neutrino sector. At proton-proton colliders like the LHC, mono-X searches, such as monojet signatures originally developed for dark matter, are equally applicable to NSIs, as replacing dark matter with neutrinos results in similar missing energy signatures. Existing LHC studies have already constrained NSI couplings to quarks through such channels \cite{Babu:2020nna, Choudhury:2018xsm, BuarqueFranzosi:2015qil, Friedland:2011za, Davidson:2011kr, Freitas:2025bgg, Lozano:2025ekx, Liu:2020emq}.
In this work, we turn our attention to neutrino-electron interactions \footnote{For insights into muon–neutrino interactions, refer to the muon collider study in Ref.~\cite{Jana:2023ogd}.}. We focus on monophoton signals and first derive limits using LEP data. We then project sensitivities at future lepton colliders such as ILC, CLIC, and FCC-ee. Furthermore, we explore the potential of probing leptonic NSIs via lepton parton distribution functions (PDFs) at the FCC. Our findings suggest that if NSIs arise from heavy mediators, lepton colliders could offer superior sensitivity compared to oscillation experiments and may help resolve existing parameter degeneracies.\\

\begin{figure}[t]
    \centering
    \includegraphics[width=0.95\linewidth]{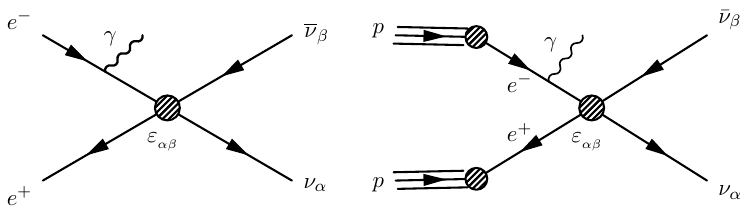}
    (a) $e^-\text{-}\,e^+$ Collider,\hspace{1.8cm}(b) $p\text{-}p$ Collider.\hspace{0.5cm}
    \caption{Representative Feynman diagrams for production of neutrino pairs via leptonic NSI at (a) electron-positron and (b) proton-proton collisions along with an ISR photon. }
    \label{fig:FeynNSI}
\end{figure}

\textbf{\emph{Neutrino NSI: From EFT to Simplified Models}.--- } 
The effects of beyond Standard Model neutrino interactions, manifesting as non-standard interactions (NSI) with matter, are suitably described within the effective field theory (EFT) framework and, at leading order, are encoded in dimension-six operators by
\begin{align}
    \mathcal{L}_{NSI} &= -2\sqrt{2}G_F \,\epsilon_{\alpha\beta} \left(\overline{\nu_\alpha}\gamma_\mu P_L\nu_\beta\right)\left(\overline{f}\gamma^\mu P_X f\right).
    \label{eq:4fNSI}
\end{align}
Here, $\alpha, \beta \in {e, \mu, \tau}$ denote the neutrino flavor indices, and $P_X \equiv P_{L,R} = \frac{1}{2}(1 \mp \gamma^5)$ are the left- and right-handed chirality projection operators. The parameter $\epsilon_{\alpha\beta}$ characterizes the strength of the NSI, while $f$ represents generic matter fields, such as electrons or up and down quarks. In this work, we focus on leptonic NSI involving neutrino interactions with electrons; therefore, in Eq.\eqref{eq:4fNSI}, we take $f \equiv e$. This class of interactions can be primarily probed at future $e^+e^-$ colliders. However, due to the presence of leptonic contributions in the proton parton distribution functions (PDFs), they may also be accessible at the LHC, as shown in Fig.\ref{fig:FeynNSI}.

In neutrino oscillation experiments, the EFT approach serves as a robust framework, as it captures the relevant phenomenology without requiring explicit details of the underlying operator structure. However, this approach becomes insufficient in high-energy collider environments, such as the ILC, where it is more appropriate to employ simplified models featuring explicit mediators. These models provide a more accurate and complete description of the interactions over a wide range of momentum transfers. While oscillation experiments allow for an agnostic treatment of the origin of NSI operators, owing to their insensitivity to the mediator mass, collider sensitivities, in contrast, are strongly dependent on the mediator’s mass and coupling structure. As a result, the EFT framework may lose its validity at the energy scales probed by colliders. In this work, we focus on the case of the future $e^+e^-$ colliders and, assuming a vector mediator, we reformulate Eq.~\eqref{eq:4fNSI} to reflect this simplified model perspective.
\begin{equation}
    \mathcal{L}_{NSI}^{Simp} = \left((g_\nu)_{\alpha\beta}\; \overline{\nu}_\alpha\gamma^\mu P_L\nu_\beta + g_e^X \overline{e}\gamma^\mu P_X e\right)\, Z^\prime_\mu
    \label{eq:NSIsimp}
\end{equation}
where $Z'$ denotes the field associated with the new force mediator of mass $M_{Z'}$, and $g_\nu$ and $g_e^X$ represent its couplings to neutrinos and electrons, respectively.

When the mass of the \( Z' \) boson is smaller than the typical momentum transfer, the interaction exhibits a resonant enhancement, whereas for sufficiently large \( M_{Z'} \), the process smoothly transitions to the EFT regime relevant to the energy scale of \( e^+e^- \) colliders. Integrating out the \( Z' \) field yields the corresponding NSI as 
\begin{equation}
    \epsilon_{\alpha\beta} = \frac{(g_\nu)_{\alpha\beta}\,g_e^V}{2\sqrt{2}G_FM_{Z^\prime}^2},
    \label{eq:NSI}
\end{equation}
where, $g^V_e = g^L_e + g^R_e$ denotes the vector current coupling. In the following section, we present a detailed discussion of the associated phenomenology at future $e^+e^-$ colliders.
\\

\textbf{\emph{Collider Analysis}.---} To explore leptonic NSI at high-energy colliders, we consider mono-$X$ signatures, where $X$ denotes a tagging particle emitted from the initial-state leptons. Our study focuses on three proposed $e^+e^-$ collider facilities: the International Linear Collider (ILC)~\cite{Behnke:2013xla}, the Compact Linear Collider (CLIC)~\cite{CLIC:2016zwp}, and the Future Circular Collider (FCC-$ee$)~\cite{FCC:2018evy}. Furthermore, motivated by the observation in Ref.~\cite{Buonocore:2020nai} that the proton exhibits a non-negligible leptonic parton density function (PDF), we also identify the LHC as a complementary platform, particularly in light of its high center-of-mass energy and luminosity, for probing leptonic NSI involving neutrinos. In all collider settings, we focus on the monophoton channel, which offers the most promising sensitivity for probing leptonic initial states. The model has been implemented in \texttt{FeynRules}~\cite{Alloul:2013bka} to facilitate subsequent simulations with Monte Carlo event generators.

\begin{table}[b]
    \centering
    \caption{\label{tab:gencuts}%
  Basic phase-space cuts applied for event generation at the lepton colliders ILC, CLIC, and FCC-$ee$, for various center-of-mass energies. }
    \begin{ruledtabular}
	\resizebox{\linewidth}{!}{
    \begin{tabular}{c|ccc}
     \cmmnt{& LEP} & ILC & CLIC & FCC-$ee$ \\
     \hline
     $\sqrt{s}$ & 1 TeV & 3 TeV & 365 GeV \\
     \hline
     $\mathcal{L}_{int}$ & 8 ab$^{-1}$ & 5 ab$^{-1}$ & 1.5 ab$^{-1}$ \\
     \hline
     \cmmnt{& \phantom{$p_T^\gamma>2$ GeV,}} & $p_T^\gamma>2$ GeV, & $p_T^\gamma>2$ GeV, & $p_T^\gamma>1$ GeV, \\
     Basic \cmmnt{& \phantom{$|\cos{\theta}|<0.9975$,}} & $|\cos{\theta}|<0.9975$, & $|\cos{\theta}|<0.9962$, & $|\cos{\theta}|<0.9962$, \\
     Cuts \cmmnt{& \phantom{$E_\gamma>1$ GeV}} & $E_\gamma>1$ GeV & $E_\gamma>1$ GeV & $E_\gamma>1$ GeV \\
    \end{tabular}
    }
    \end{ruledtabular}
    
\end{table}

\emph{Lepton Colliders---} We perform event generation at lepton colliders using \texttt{Whizard}~\cite{Kilian:2007gr}, applying basic cuts corresponding to the highest upgrade configurations, as detailed in Table~\ref{tab:gencuts}. The presence of missing energy in the signal makes it susceptible to various Standard Model (SM) backgrounds. The most dominant contribution comes from neutrino pair production accompanied by an initial-state radiation (ISR) photon. An additional irreducible background arises from radiative Bhabha scattering when final-state electrons escape detection. This process exhibits collinear divergence, which we control by introducing additional kinematic cuts, following the method described in Ref.~\cite{Habermehl:2018yul},
\begin{align}
    &M_{e^\pm_{in},e^\pm_{out}} < -2 \text{ GeV, } &M_{e^\pm_{out},e^\pm_{out}} > 2 \text{ GeV, } \nonumber\\
    &M_{e^\pm_{out},\gamma_i}\;\, > 2 \text{ GeV, } &M_{e^\pm_{out},\gamma} > 4 \text{ GeV, }
\end{align}
where $\gamma_i$ are the ME (matrix element) photons and $\gamma$ denotes the signal photon. We also implement the ME-ISR merging to remove double counting due to multiple photon emission~\cite{Kalinowski:2020lhp}. For that, we have used two independent variables $q_\pm$, defined as:
\begin{align}
    q_+\equiv \sqrt{2E_{cm}E_\gamma}\sin{\frac{\theta}{2}} \nonumber\\
    q_-\equiv \sqrt{2E_{cm}E_\gamma}\cos{\frac{\theta}{2}}.
\end{align}
Here, $E_{cm} = \sqrt{s}$ denotes the center-of-mass energy, $E_\gamma$ is the photon energy, and $\theta_\gamma$ represents the photon polar angle. Events are discarded if they contain ME photons with $q_\pm < 1$~GeV and $E_\gamma < 1$~GeV, or ISR photons with $q_\pm > 1$~GeV and $E_\gamma > 1$~GeV.

Detector effects are simulated using \texttt{Delphes3}~\cite{deFavereau:2013fsa} package, with the \texttt{ILCgen}, \texttt{CLICdet\_Stage3\_fcal}, and \texttt{IDEA} configuration cards for ILC, CLIC, and FCC-$ee$, respectively.  For all lepton collider scenarios, photons are identified with $E_\gamma > 5$~GeV and pseudorapidity  $|\eta| < 2.8$, electrons with $p_T > 2$~GeV and $|\eta| < 2.8$, and muons and jets with $p_T > 5$~GeV and $|\eta| < 2.8$. Jets are reconstructed using the \texttt{FastJet3}~\cite{Cacciari:2011ma} package with the anti-$k_T$ algorithm~\cite{Cacciari:2008gp}, applying cone radius parameter $R = 1.0$ and a minimum transverse momentum threshold of $p_T^{\text{min}} > 5$~GeV.

For event selection at the ILC, we require events to contain at least one photon with energy $E_\gamma > 5$~GeV and $|\eta_\gamma| < 2.8$. To suppress background contributions, events containing charged leptons or jets are vetoed. In addition, we impose a veto on energy depositions in the very forward electromagnetic calorimeter (BeamCal)~\cite{Abramowicz:2010bg}, where such depositions are reconstructed as BeamCal photons. Events containing BeamCal photons are excluded. These vetoes are particularly effective in reducing the radiative Bhabha background. To further suppress background, we apply an angular cut of $|\cos{\theta}| < 0.95$. 

The photon energy distribution of the signal depends on the mass of the mediator. For $M_{Z^\prime} < \sqrt{s}$, the signal exhibits a resonance-return peak at $E_\gamma = \frac{s - M_{Z^\prime}^2}{2\sqrt{s}}$, analogous to the $Z$-return peak~\cite{KumarRai:2003kk}. 

For low mediator masses, especially near the $Z$-boson mass, this peak overlaps with that of the background, making photon energy cuts ineffective. In contrast, for larger mediator masses, the high-energy tail of the photon spectrum becomes background-dominated, allowing for additional background suppression via an upper cut on photon energy. In the case where $M_{Z^\prime} \ge \sqrt{s}$, no resonant enhancement is observed, and a fixed cut of $E_\gamma < 350$~GeV is applied.

\begin{figure*}[t]
    \centering
    \includegraphics[width=0.65\linewidth]{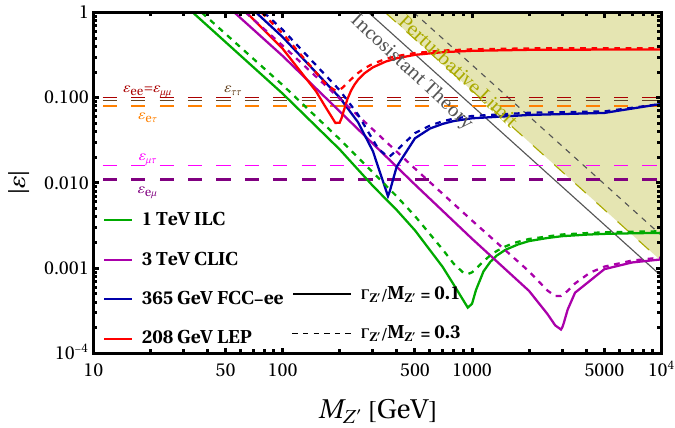}
    \caption{The projected 3$\sigma$ sensitivity to neutrino NSI is shown as a function of the mediator mass for various $e^+e^-$ colliders operating at their highest proposed upgrade configurations. Solid and dashed lines correspond to two different assumptions for the mediator width, specifically $\Gamma_{Z^\prime} = 0.1M_{Z^\prime}$ and $\Gamma_{Z^\prime} = 0.3M_{Z^\prime}$, respectively. The shaded region, along with the gray solid and dashed lines, indicates the parameter space excluded by perturbativity constraints and theoretical inconsistency arising from decay width inequality constraints. Horizontal dashed lines represent current bounds on $\epsilon_{\alpha\beta}$ from neutrino oscillation data for different flavor combinations.}
    \label{fig:limit}
\end{figure*}

At ILC, both electron and positron beams can be polarized. Among the available configurations, the polarization choice $P(e^-, e^+) = (-80\%, +30\%)$ yields the most effective suppression of SM backgrounds~\cite{Habermehl:2020njb,Kundu:2021cmo}. However, to mitigate systematic uncertainties, we adopt the H20 scenario of mixed polarization  configurations~\cite{Barklow:2015tja}. The signal significance is evaluated using the following expression:
\begin{equation}
    Sig=\frac{S}{\sqrt{\sum_i(B_i+\epsilon_i^2 B_i^2)}}
    \label{eq:significance}
\end{equation}
Here, $S$ represents the number of signal events, and $B_i$ denotes the number of events from the $i$-th background, with $\epsilon_i$ corresponding to its systematic error. We consider systematic uncertainties of 1\% for the neutrino background and 0.2\% for the Bhabha background~\cite{Kalinowski:2020lhp, Habermehl:2018yul, Blaising:2021vhh}. The significance is calculated for each polarization configuration and combined to yield the effective significance for the H20 scenario as described by the following formula:
\begin{equation}
    Sig_{H20}=\sqrt{\sum_iSig_i^2}.
    \label{eq:H20}
\end{equation}
Using Eq.~\eqref{eq:H20}, we determine the $3\sigma$ sensitivity limits on the NSI parameters, $\epsilon_{\alpha\beta}$, for various $Z^\prime$ masses assuming vector couplings to electrons. The resulting constraints are illustrated in Fig.~\ref{fig:limit}, where the green solid and dashed lines correspond to mediator widths of 10\% and 30\% of $M_{Z^\prime}$, respectively.

For the CLIC analysis, we select events containing at least one photon with $E_\gamma > 5$~GeV and $|\eta_\gamma| < 2.8$, while rejecting those that include additional charged leptons or jets. A forward electromagnetic calorimeter is also part of the CLIC detector design, where energy deposits may be reconstructed as photons; events with such forward photons are vetoed to suppress background. Background reduction strategies mirror those employed in the ILC case, including the application of an angular cut $|\cos{\theta_\gamma}| < 0.95$ and a photon energy threshold tailored to the mediator mass. Since positron beam polarization is not available at CLIC, we consider electron beam polarization only, with an 80\% polarization in both left- and right-helicity modes. The signal significance is computed using Eq.~\eqref{eq:significance}, and combined across polarizations to derive the $3\sigma$ sensitivity on the NSI parameter $\epsilon_{\alpha\beta}$. The corresponding sensitivities are shown in Figure~\ref{fig:limit}, where magenta solid and dashed lines indicate results for $Z^\prime$ widths of 10\% and 30\% of $M_{Z^\prime}$, respectively.

For the FCC-$ee$ analysis, events are selected by requiring at least one photon with $E_\gamma > 5$~GeV and pseudorapidity $|\eta_\gamma| < 2.8$, while excluding events containing charged leptons or jets. Backgrounds are suppressed using the same analysis strategy as in previous cases, applying the angular cut $|\cos{\theta_\gamma}| < 0.95$ and a photon energy threshold dependent on the mass of the $Z^\prime$ mediator. The resulting $3\sigma$ sensitivity bounds, computed using Eq.~\eqref{eq:significance}, are shown in Fig.~\ref{fig:limit}. Note that for the FCC-$ee$, the beam polarization feature is absent; therefore, we have worked only with unpolarised beams. The blue solid and dashed curves correspond to mediator widths of 10\% and 30\% of $M_{Z^\prime}$, respectively.

In all cases, the highest sensitivity is achieved for resonant production of the vector mediator $Z^\prime$ when $M_{Z^\prime} = \sqrt{s}$ with a width $\Gamma_{Z^\prime} = 0.1 \times M_{Z^\prime}$. For a larger width, the sensitivity is slightly reduced, showing a broader dip as expected. The red lines represent constraints from the LEP monophoton search, with the strongest limit coming from the L3 detector~\cite{L3:2003yon}. We use the following relation to estimate the 95\% confidence level bound on the NSI parameter~\cite{Berezhiani:2001rs}:
\begin{equation}
    \sigma^{NSI}\le |\sigma^{obs}+3\delta\sigma^{obs}-\sigma^{exp}|
\end{equation}
where $\sigma^{NSI}$ denotes the cross-section due to neutrino NSI, $\sigma^{obs} \pm \delta\sigma^{obs}$ is the observed cross-section measured by L3, and $\sigma^{exp}$ represents the Standard Model expected cross-section. To estimate $\sigma^{NSI}$, we replicated the event topology for the multi-photon channel as detailed in Ref.~\cite{L3:2003yon}.

Before presenting further results, it is important to discuss the validity of the EFT treatment for neutrino NSI at colliders. A meaningful constraint on the NSI parameter can be obtained by examining the consistency of the EFT with an underlying simplified model. In our setup, the total decay width of the mediator $Z^\prime$ must be at least as large as the sum of its partial widths into electrons and neutrinos:
\begin{equation}
    \Gamma_{Z^\prime} \geq \Gamma_{Z^\prime \to \bar{\nu} \nu} + \Gamma_{Z^\prime \to e^+ e^-}.
\end{equation}
\begin{figure}[t]
    \centering
    \includegraphics[width=0.9\linewidth]{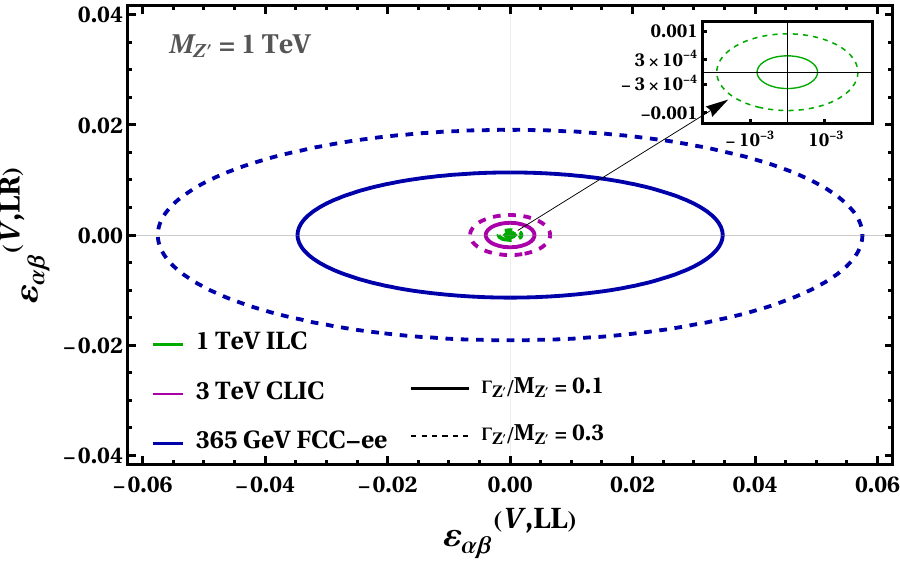}
    \caption{The sensitivity reach in the $(\epsilon_{\alpha\beta}^{V,LL}, \epsilon_{\alpha\beta}^{V,LR})$ parameter space is presented for a benchmark mediator mass of 1~TeV, taking into account the three proposed future $e^+e^-$ colliders. The solid and dashed contours correspond to two representative choices of the mediator width, as described in the analysis.
}
    \label{fig:LRcontour}
\end{figure}

This leads to a constraint on the partial width for a single neutrino flavor,
\begin{equation}
    \Gamma_{Z^\prime} \geq \frac{M_{Z^\prime}}{24\pi} \left[g_\nu^2 + (g_e^V)^2\right] \geq \frac{M_{Z^\prime}}{24\pi} \cdot 2\sqrt{2} g_\nu g_e^V.
\end{equation}
Consequently, this implies an upper bound on the NSI parameter $\epsilon$, expressed as:
\begin{equation}
    |\epsilon| \leq \frac{3\pi}{G_F M_{Z^\prime}^2} \cdot \frac{\Gamma_{Z^\prime}}{M_{Z^\prime}}.
\end{equation}
This theoretical constraint is shown in Fig.~\ref{fig:limit} with gray solid and dashed lines corresponding to two different width-to-mass ratios, namely $\Gamma_{Z^\prime}/M_{Z^\prime} = 0.1$ and $0.3$, respectively. A smaller width leads to a more stringent bound on $\epsilon$. We also ensure that both $g_\nu$ and $g_e^V$ remain within the perturbative regime.

We have also included complementary constraints from neutrino oscillation experiments, where the flavor structure of the NSIs and the choice of oscillation channels play a significant role. In Fig.~\ref{fig:limit}, we present the sensitivity reach of $|\epsilon_{\alpha\beta}|$ for all flavor combinations, based on a global fit to oscillation data~\cite{Coloma:2023ixt}. The limits range from 0.01 to 0.1, with the strongest constraints arising from flavor-violating NSIs. Our findings indicate that the sizeable flavor-changing NSIs, which have been proposed as a possible explanation for the observed discrepancies in the extracted values of the CP-violating phase $\delta_{\mathrm{CP}}$ between NO$\nu$A and T2K \cite{Denton:2020uda, Chatterjee:2020kkm}, can be entirely ruled out by future monophoton searches at lepton colliders. 

In the SM, neutrinos belong to the same $SU(2)_L$ multiplet as charged leptons. Consequently, NSIs involving different neutrino flavors generally induce sizable charged lepton flavor violation, which is tightly constrained. However, these bounds may be relaxed if the mediator is light or if the NSIs are generated by higher-dimensional operators (\emph{e.g.}, dimension-8). In contrast, flavor-diagonal NSIs can remain sizable in several neutrino mass models. Fig.~\ref{fig:limit} also illustrates that lepton colliders offer superior sensitivity compared to oscillation experiments when the mediator mass exceeds the electroweak scale. This highlights a notable complementarity: oscillation experiments are sensitive to neutrino flavor but not the mediator mass, though they suffer from parameter degeneracies where NSI effects may cancel out. Lepton colliders, on the other hand, are sensitive to the mediator mass and are not affected by neutrino flavor, enabling them to resolve such degeneracies. For completeness, Fig.~\ref{fig:LRcontour} shows the NSI constarints in the $(\epsilon_{\alpha\beta}^{V,LL}, \epsilon_{\alpha\beta}^{V,LR})$ plane for a 1~TeV mediator. At lepton colliders, sensitivity to $\epsilon_{\alpha\beta}^{V,LL}$ is generally stronger than to $\epsilon_{\alpha\beta}^{V,LR}$, though both are comparable when $\epsilon_{\alpha\beta}^{V,LL} = \epsilon_{\alpha\beta}^{V,LR}$.
 \\
\begin{figure}[t]
    \centering
    \includegraphics[width=0.9\linewidth]{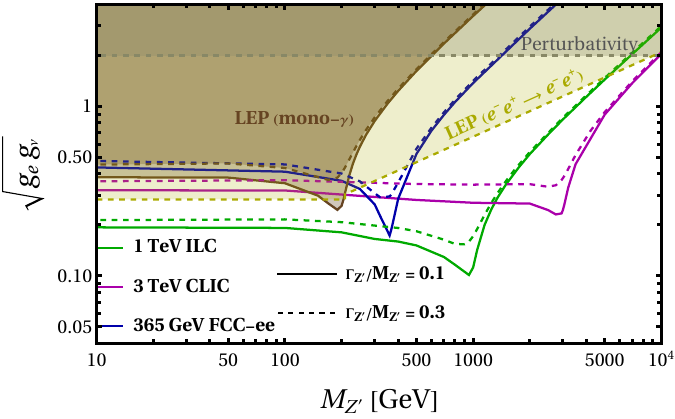}
    \caption{The projected $3\sigma$ sensitivity reach of $\sqrt{g_e g_\nu}$ as a function of the mediator mass is shown for future lepton colliders. For comparison, we include existing constraints from the LEP monophoton search (brown shaded region) and the LEP $e^-e^+ \to e^-e^+$ scattering data (yellow shaded region), assuming the neutrino coupling to the vector mediator remains within the perturbative regime. The gray shaded region indicates the parameter space excluded by the perturbativity bounds.}
    \label{fig:gegnu}
\end{figure}

\begin{figure*}[t]
    \centering
    \includegraphics[width=0.9\linewidth]{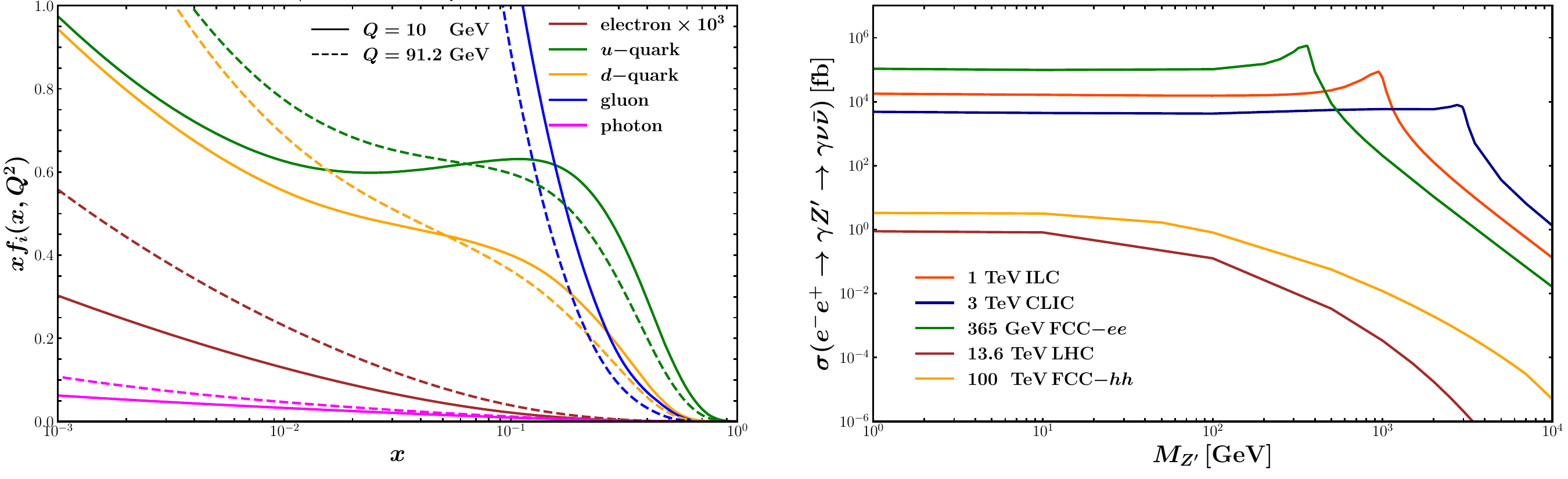}
    \caption{ (\emph{Left}) Parton distribution functions for different constituents in proton. (\emph{Right}) Cross-sections at different colliders as a function of the cut-off scale ($Z^\prime$ mass)}
    \label{fig:PDF}
\end{figure*}

Having established the potential of future lepton colliders to probe neutrino NSI across different energy scales by looking at monophoton signal, we now focus on constraints from other relevant collider searches. Since left-handed neutrinos and charged leptons belong to the same electroweak $SU(2)_L$ multiplet, any UV-complete realization typically leads to couplings of the mediator with both neutrinos and charged leptons. As a result, a portion of the NSI parameter space that is accessible through neutrino signals at $e^+e^-$ colliders may also be constrained by charged lepton pair-production processes, such as $e^+e^- \to e^+e^-$, due to the mediator's coupling to electrons. Fig.~\ref{fig:gegnu} illustrates the projected $3\sigma$ sensitivity to the product of $Z^\prime$ couplings to electrons and neutrinos, $g_e  g_\nu$, as a function of the $Z^\prime$ mass. The mass range is selected such that the coupling to neutrinos, $g_\nu$, can be maximized up to the perturbative limit. In our analysis, we allow $g_\nu$ to reach up to 2, given that current experimental bounds do not constrain this coupling in the chosen mass window. The electron coupling, $g_e$, is then adjusted within allowed values to generate detectable NSI while satisfying experimental constraints. The gray shaded region in the plot represents the parameter space excluded by the perturbativity condition. The brown shaded region is constrained by LEP monophoton searches, recast from Ref.~\cite{OPAL:2003kcu}. We see that the LEP constraint from the monophoton channel is strongest at the resonance point and gradually weakens at higher mass regions. In the higher mass region, however, the $e^+e^-\to e^+e^-$ gives comparatively stronger constraint. The yellow shaded area corresponds to limits from fermion pair production measurements by the OPAL Collaboration at LEP2. In the region above the center-of-mass energy, 207 GeV, where the EFT validity is ensured, we put a constraint at $g_e>2.2\times10^{-4}M_{Z^\prime}$. In the low mass region, we have put a flat constraint of $0.04$~\cite{Buckley:2011vc}. From the plot, it is evident that the ILC offers the best sensitivity near the resonant production of the vector mediator. Moreover, we observe that part of the NSI parameter space that can be probed via monophoton signal at future lepton colliders is partially constrained by existing $e^+e^- \to e^+e^-$ measurements.

At hadron colliders such as the LHC, the incoming high-energy proton beams can radiate photons, which in turn may produce energetic leptons through QED interactions. Although leptons are not valence constituents of the proton, they can still initiate processes involving large momentum transfer, enabling the exploration of higher-order perturbative effects. In Fig.~\ref{fig:PDF}, we illustrate the evolution of various parton distribution functions (PDFs) with respect to the momentum fraction, $x$. While the electron density in the proton is indeed much lower than that of colored partons, lepton-initiated processes—though rare—offer an interesting window into otherwise inaccessible regions of parameter space.
We present in Fig.~\ref{fig:PDF} the cross-sections for the process $e^-e^+ \to Z^\prime \to \nu_\alpha \overline{\nu}_\beta$ at the LHC and compare them with those at future lepton colliders. Although the LHC environment is characterized by substantial backgrounds, especially from processes such as $Z+\gamma$ and $W+\gamma$, this study highlights the challenge and motivates the need for advanced search strategies.

\textbf{\emph{Conclusions}.--- }We conduct a comprehensive study on the prospects of future lepton colliders—specifically the ILC, CLIC, and FCC-ee—for probing leptonic non-standard interactions (NSIs), with a focus on scenarios mediated by heavy particles on or above the electroweak scale. These colliders offer substantially enhanced sensitivity compared to conventional neutrino oscillation experiments, particularly across a wide range of mediator masses. Our analysis demonstrates that CLIC, operating at a center-of-mass energy of 3 TeV, achieves the highest sensitivity, constraining NSI effects down to $0.019\%$. The ILC reaches sensitivities at the $0.034\%$ level, while the FCC-ee provides bounds at approximately $0.7\%$. Additionally, we investigate the complementary capabilities of future proton-proton colliders, such as the FCC-$hh$ with $\sqrt{s}=100$ TeV, to probe NSIs through lepton parton distribution functions (PDFs), thereby offering a novel pathway for indirect searches of physics beyond the Standard Model. Our results highlight the crucial role of collider data in resolving parameter degeneracies that hinder the precise interpretation of oscillation measurements. Notably, scenarios involving sizeable NSI strengths—proposed as potential explanations for anomalies such as the tension between T2K and NO$\nu$A observations—can be stringently tested and decisively excluded.
\vspace{0.1in}
\begin{acknowledgments}
{\textbf {\textit Acknowledgments.---}} SJ, SK and SKR would like to acknowledge the support from Department of Atomic Energy (DAE), Government of India for the Regional Centre for Accelerator based Particle Physics (RECAPP), Harish-Chandra Research Institute.
\end{acknowledgments}

\bibliographystyle{utphys}
\bibliography{reference}

\providecommand{\href}[2]{#2}\begingroup\raggedright\begin{thebibliography}{10}

\bibitem{Proceedings:2019qno}
\href{http://dx.doi.org/10.21468/SciPostPhysProc.2.001}{{\em {Neutrino Non-Standard Interactions: A Status Report}}}, vol.~2.
\newblock 2019.
\newblock \href{http://arxiv.org/abs/1907.00991}{{\ttfamily arXiv:1907.00991 [hep-ph]}}.

\bibitem{Babu:2019mfe}
K.~S. Babu, P.~S.~B. Dev, S.~Jana, and A.~Thapa, ``{Non-Standard Interactions in Radiative Neutrino Mass Models},'' \href{http://dx.doi.org/10.1007/JHEP03(2020)006}{{\em JHEP} {\bfseries 03} (2020) 006}, \href{http://arxiv.org/abs/1907.09498}{{\ttfamily arXiv:1907.09498 [hep-ph]}}.

\bibitem{Wolfenstein:1977ue}
L.~Wolfenstein, ``{Neutrino Oscillations in Matter},'' \href{http://dx.doi.org/10.1103/PhysRevD.17.2369}{{\em Phys. Rev. D} {\bfseries 17} (1978) 2369--2374}.

\bibitem{Miranda:2004nb}
O.~G. Miranda, M.~A. Tortola, and J.~W.~F. Valle, ``{Are solar neutrino oscillations robust?},'' \href{http://dx.doi.org/10.1088/1126-6708/2006/10/008}{{\em JHEP} {\bfseries 10} (2006) 008}, \href{http://arxiv.org/abs/hep-ph/0406280}{{\ttfamily arXiv:hep-ph/0406280}}.

\bibitem{Maltoni:2015kca}
M.~Maltoni and A.~Y. Smirnov, ``{Solar neutrinos and neutrino physics},'' \href{http://dx.doi.org/10.1140/epja/i2016-16087-0}{{\em Eur. Phys. J. A} {\bfseries 52} no.~4, (2016) 87}, \href{http://arxiv.org/abs/1507.05287}{{\ttfamily arXiv:1507.05287 [hep-ph]}}.

\bibitem{Denton:2020uda}
P.~B. Denton, J.~Gehrlein, and R.~Pestes, ``{$CP$ -Violating Neutrino Nonstandard Interactions in Long-Baseline-Accelerator Data},'' \href{http://dx.doi.org/10.1103/PhysRevLett.126.051801}{{\em Phys. Rev. Lett.} {\bfseries 126} no.~5, (2021) 051801}, \href{http://arxiv.org/abs/2008.01110}{{\ttfamily arXiv:2008.01110 [hep-ph]}}.

\bibitem{Chatterjee:2020kkm}
S.~S. Chatterjee and A.~Palazzo, ``{Nonstandard Neutrino Interactions as a Solution to the $NO\nu A$ and T2K Discrepancy},'' \href{http://dx.doi.org/10.1103/PhysRevLett.126.051802}{{\em Phys. Rev. Lett.} {\bfseries 126} no.~5, (2021) 051802}, \href{http://arxiv.org/abs/2008.04161}{{\ttfamily arXiv:2008.04161 [hep-ph]}}.

\bibitem{Babu:2020nna}
K.~S. Babu, D.~Gon\c{c}alves, S.~Jana, and P.~A.~N. Machado, ``{Neutrino Non-Standard Interactions: Complementarity Between LHC and Oscillation Experiments},'' \href{http://dx.doi.org/10.1016/j.physletb.2021.136131}{{\em Phys. Lett. B} {\bfseries 815} (2021) 136131}, \href{http://arxiv.org/abs/2003.03383}{{\ttfamily arXiv:2003.03383 [hep-ph]}}.

\bibitem{Choudhury:2018xsm}
D.~Choudhury, K.~Ghosh, and S.~Niyogi, ``{Probing nonstandard neutrino interactions at the LHC Run II},'' \href{http://dx.doi.org/10.1016/j.physletb.2018.07.053}{{\em Phys. Lett. B} {\bfseries 784} (2018) 248--254}, \href{http://arxiv.org/abs/1801.01513}{{\ttfamily arXiv:1801.01513 [hep-ph]}}.

\bibitem{BuarqueFranzosi:2015qil}
D.~Buarque~Franzosi, M.~T. Frandsen, and I.~M. Shoemaker, ``{New or $\nu$ missing energy: Discriminating dark matter from neutrino interactions at the LHC},'' \href{http://dx.doi.org/10.1103/PhysRevD.93.095001}{{\em Phys. Rev. D} {\bfseries 93} no.~9, (2016) 095001}, \href{http://arxiv.org/abs/1507.07574}{{\ttfamily arXiv:1507.07574 [hep-ph]}}.

\bibitem{Friedland:2011za}
A.~Friedland, M.~L. Graesser, I.~M. Shoemaker, and L.~Vecchi, ``{Probing Nonstandard Standard Model Backgrounds with LHC Monojets},'' \href{http://dx.doi.org/10.1016/j.physletb.2012.06.078}{{\em Phys. Lett. B} {\bfseries 714} (2012) 267--275}, \href{http://arxiv.org/abs/1111.5331}{{\ttfamily arXiv:1111.5331 [hep-ph]}}.

\bibitem{Davidson:2011kr}
S.~Davidson and V.~Sanz, ``{Non-Standard Neutrino Interactions at Colliders},'' \href{http://dx.doi.org/10.1103/PhysRevD.84.113011}{{\em Phys. Rev. D} {\bfseries 84} (2011) 113011}, \href{http://arxiv.org/abs/1108.5320}{{\ttfamily arXiv:1108.5320 [hep-ph]}}.

\bibitem{Freitas:2025bgg}
A.~Freitas and M.~Low, ``{Non-Standard Neutrino Interactions at Neutrino Experiments and Colliders},'' \href{http://arxiv.org/abs/2505.01401}{{\ttfamily arXiv:2505.01401 [hep-ph]}}.

\bibitem{Lozano:2025ekx}
V.~M. Lozano, G.~Sanchez~Garcia, and A.~Terrones, ``{Neutrino Non-Standard Interactions: Confronting COHERENT and LHC},'' \href{http://arxiv.org/abs/2503.11766}{{\ttfamily arXiv:2503.11766 [hep-ph]}}.

\bibitem{Liu:2020emq}
D.~Liu, C.~Sun, and J.~Gao, ``{Constraints on neutrino non-standard interactions from LHC data with large missing transverse momentum},'' \href{http://dx.doi.org/10.1007/JHEP02(2021)033}{{\em JHEP} {\bfseries 02} (2021) 033}, \href{http://arxiv.org/abs/2009.06668}{{\ttfamily arXiv:2009.06668 [hep-ph]}}.

\bibitem{Note1}
For insights into muon–neutrino interactions, refer to the muon collider study in Ref.~\cite {Jana:2023ogd}.

\bibitem{Behnke:2013xla}
``{The International Linear Collider Technical Design Report - Volume 1: Executive Summary},''  (6, 2013) , \href{http://arxiv.org/abs/1306.6327}{{\ttfamily arXiv:1306.6327 [physics.acc-ph]}}.

\bibitem{CLIC:2016zwp}
{\bfseries CLIC, CLICdp} Collaboration, M.~J. Boland {\em et~al.}, ``{Updated baseline for a staged Compact Linear Collider},'' \href{http://dx.doi.org/10.5170/CERN-2016-004}{ (8, 2016) }, \href{http://arxiv.org/abs/1608.07537}{{\ttfamily arXiv:1608.07537 [physics.acc-ph]}}.

\bibitem{FCC:2018evy}
{\bfseries FCC} Collaboration, A.~Abada {\em et~al.}, ``{FCC-ee: The Lepton Collider}: {Future Circular Collider Conceptual Design Report Volume 2},'' \href{http://dx.doi.org/10.1140/epjst/e2019-900045-4}{{\em Eur. Phys. J. ST} {\bfseries 228} no.~2, (2019) 261--623}.

\bibitem{Buonocore:2020nai}
L.~Buonocore, P.~Nason, F.~Tramontano, and G.~Zanderighi, ``{Leptons in the proton},'' \href{http://dx.doi.org/10.1007/JHEP08(2020)019}{{\em JHEP} {\bfseries 08} no.~08, (2020) 019}, \href{http://arxiv.org/abs/2005.06477}{{\ttfamily arXiv:2005.06477 [hep-ph]}}.

\bibitem{Alloul:2013bka}
A.~Alloul, N.~D. Christensen, C.~Degrande, C.~Duhr, and B.~Fuks, ``{FeynRules 2.0 - A complete toolbox for tree-level phenomenology},'' \href{http://dx.doi.org/10.1016/j.cpc.2014.04.012}{{\em Comput. Phys. Commun.} {\bfseries 185} (2014) 2250--2300}, \href{http://arxiv.org/abs/1310.1921}{{\ttfamily arXiv:1310.1921 [hep-ph]}}.

\bibitem{Kilian:2007gr}
W.~Kilian, T.~Ohl, and J.~Reuter, ``{WHIZARD: Simulating Multi-Particle Processes at LHC and ILC},'' \href{http://dx.doi.org/10.1140/epjc/s10052-011-1742-y}{{\em Eur. Phys. J. C} {\bfseries 71} (2011) 1742}, \href{http://arxiv.org/abs/0708.4233}{{\ttfamily arXiv:0708.4233 [hep-ph]}}.

\bibitem{Habermehl:2018yul}
M.~Habermehl, \href{http://dx.doi.org/10.3204/PUBDB-2018-05723}{{\em {Dark Matter at the International Linear Collider}}}.
\newblock PhD thesis, Hamburg U., Hamburg, 2018.

\bibitem{Kalinowski:2020lhp}
J.~Kalinowski, W.~Kotlarski, P.~Sopicki, and A.~F. Zarnecki, ``{Simulating hard photon production with WHIZARD},'' \href{http://dx.doi.org/10.1140/epjc/s10052-020-8149-6}{{\em Eur. Phys. J. C} {\bfseries 80} no.~7, (2020) 634}, \href{http://arxiv.org/abs/2004.14486}{{\ttfamily arXiv:2004.14486 [hep-ph]}}.

\bibitem{deFavereau:2013fsa}
{\bfseries DELPHES 3} Collaboration, J.~de~Favereau, C.~Delaere, P.~Demin, A.~Giammanco, V.~Lema\^\i{}tre, A.~Mertens, and M.~Selvaggi, ``{DELPHES 3, A modular framework for fast simulation of a generic collider experiment},'' \href{http://dx.doi.org/10.1007/JHEP02(2014)057}{{\em JHEP} {\bfseries 02} (2014) 057}, \href{http://arxiv.org/abs/1307.6346}{{\ttfamily arXiv:1307.6346 [hep-ex]}}.

\bibitem{Cacciari:2011ma}
M.~Cacciari, G.~P. Salam, and G.~Soyez, ``{FastJet User Manual},'' \href{http://dx.doi.org/10.1140/epjc/s10052-012-1896-2}{{\em Eur. Phys. J. C} {\bfseries 72} (2012) 1896}, \href{http://arxiv.org/abs/1111.6097}{{\ttfamily arXiv:1111.6097 [hep-ph]}}.

\bibitem{Cacciari:2008gp}
M.~Cacciari, G.~P. Salam, and G.~Soyez, ``{The anti-$k_t$ jet clustering algorithm},'' \href{http://dx.doi.org/10.1088/1126-6708/2008/04/063}{{\em JHEP} {\bfseries 04} (2008) 063}, \href{http://arxiv.org/abs/0802.1189}{{\ttfamily arXiv:0802.1189 [hep-ph]}}.

\bibitem{Abramowicz:2010bg}
H.~Abramowicz {\em et~al.}, ``{Forward Instrumentation for ILC Detectors},'' \href{http://dx.doi.org/10.1088/1748-0221/5/12/P12002}{{\em JINST} {\bfseries 5} (2010) P12002}, \href{http://arxiv.org/abs/1009.2433}{{\ttfamily arXiv:1009.2433 [physics.ins-det]}}.

\bibitem{KumarRai:2003kk}
S.~Kumar~Rai and S.~Raychaudhuri, ``{Single photon signals for warped quantum gravity at a linear e+ e- collider},'' \href{http://dx.doi.org/10.1088/1126-6708/2003/10/020}{{\em JHEP} {\bfseries 10} (2003) 020}, \href{http://arxiv.org/abs/hep-ph/0307096}{{\ttfamily arXiv:hep-ph/0307096}}.

\bibitem{Habermehl:2020njb}
M.~Habermehl, M.~Berggren, and J.~List, ``{WIMP Dark Matter at the International Linear Collider},'' \href{http://dx.doi.org/10.1103/PhysRevD.101.075053}{{\em Phys. Rev. D} {\bfseries 101} no.~7, (2020) 075053}, \href{http://arxiv.org/abs/2001.03011}{{\ttfamily arXiv:2001.03011 [hep-ex]}}.

\bibitem{Kundu:2021cmo}
S.~Kundu, A.~Guha, P.~K. Das, and P.~S.~B. Dev, ``{EFT analysis of leptophilic dark matter at future electron-positron colliders in the mono-photon and mono-Z channels},'' \href{http://dx.doi.org/10.1103/PhysRevD.107.015003}{{\em Phys. Rev. D} {\bfseries 107} no.~1, (2023) 015003}, \href{http://arxiv.org/abs/2110.06903}{{\ttfamily arXiv:2110.06903 [hep-ph]}}.

\bibitem{Barklow:2015tja}
T.~Barklow, J.~Brau, K.~Fujii, J.~Gao, J.~List, N.~Walker, and K.~Yokoya, ``{ILC Operating Scenarios},''  (6, 2015) , \href{http://arxiv.org/abs/1506.07830}{{\ttfamily arXiv:1506.07830 [hep-ex]}}.

\bibitem{Blaising:2021vhh}
{\bfseries CLICdp} Collaboration, J.-J. Blaising, P.~Roloff, A.~Sailer, and U.~Schnoor, ``{Physics performance for Dark Matter searches at $\sqrt{s}=$ 3 TeV at CLIC using mono-photons and polarised beams},''  (3, 2021) , \href{http://arxiv.org/abs/2103.06006}{{\ttfamily arXiv:2103.06006 [hep-ex]}}.

\bibitem{L3:2003yon}
{\bfseries L3} Collaboration, P.~Achard {\em et~al.}, ``{Single photon and multiphoton events with missing energy in $e^{+} e^{-}$ collisions at LEP},'' \href{http://dx.doi.org/10.1016/j.physletb.2004.01.010}{{\em Phys. Lett. B} {\bfseries 587} (2004) 16--32}, \href{http://arxiv.org/abs/hep-ex/0402002}{{\ttfamily arXiv:hep-ex/0402002}}.

\bibitem{Berezhiani:2001rs}
Z.~Berezhiani and A.~Rossi, ``{Limits on the nonstandard interactions of neutrinos from e+ e- colliders},'' \href{http://dx.doi.org/10.1016/S0370-2693(02)01767-7}{{\em Phys. Lett. B} {\bfseries 535} (2002) 207--218}, \href{http://arxiv.org/abs/hep-ph/0111137}{{\ttfamily arXiv:hep-ph/0111137}}.

\bibitem{Coloma:2023ixt}
P.~Coloma, M.~C. Gonzalez-Garcia, M.~Maltoni, J.~a.~P. Pinheiro, and S.~Urrea, ``{Global constraints on non-standard neutrino interactions with quarks and electrons},'' \href{http://dx.doi.org/10.1007/JHEP08(2023)032}{{\em JHEP} {\bfseries 08} (2023) 032}, \href{http://arxiv.org/abs/2305.07698}{{\ttfamily arXiv:2305.07698 [hep-ph]}}.

\bibitem{OPAL:2003kcu}
{\bfseries OPAL} Collaboration, G.~Abbiendi {\em et~al.}, ``{Tests of the standard model and constraints on new physics from measurements of fermion pair production at 189-GeV to 209-GeV at LEP},'' \href{http://dx.doi.org/10.1140/epjc/s2004-01595-9}{{\em Eur. Phys. J. C} {\bfseries 33} (2004) 173--212}, \href{http://arxiv.org/abs/hep-ex/0309053}{{\ttfamily arXiv:hep-ex/0309053}}.

\bibitem{Buckley:2011vc}
M.~R. Buckley, D.~Hooper, J.~Kopp, and E.~Neil, ``{Light Z' Bosons at the Tevatron},'' \href{http://dx.doi.org/10.1103/PhysRevD.83.115013}{{\em Phys. Rev. D} {\bfseries 83} (2011) 115013}, \href{http://arxiv.org/abs/1103.6035}{{\ttfamily arXiv:1103.6035 [hep-ph]}}.

\bibitem{Jana:2023ogd}
S.~Jana and S.~Klett, ``{Muonic force and nonstandard neutrino interactions at muon colliders},'' \href{http://dx.doi.org/10.1103/PhysRevD.110.095011}{{\em Phys. Rev. D} {\bfseries 110} no.~9, (2024) 095011}, \href{http://arxiv.org/abs/2308.07375}{{\ttfamily arXiv:2308.07375 [hep-ph]}}.

\end{thebibliography}\endgroup
\appendix


\end{document}